\documentclass[9pt,twocolumn,twoside]{pnas-new}

\templatetype{pnasresearcharticle} 

\usepackage{amssymb,amsfonts,amsmath}
\usepackage{siunitx}
\usepackage{float}
\usepackage[subrefformat=parens,labelformat=parens]{subfig}
\usepackage{color}

\DeclareSIUnit\molar{M}

\newcommand{\Anum}{{\tilde A}}
\newcommand{\Tnum}{{\tilde T}}
\newcommand{\Snum}{{\tilde S}}

\newcommand{\revised}[1]{ #1}

\title{Controlling Uncertainty in Aptamer Selection}

\author[a,b]{Fabian Spill}
\author[c]{Zohar B. Weinstein} 
\author[b]{Nga Ho}
\author[b]{Atena Irani Shemirani}
\author[b,1]{Darash Desai}
\author[b,d,1]{Muhammad H. Zaman}
\affil[a]{Department of Mechanical Engineering, Massachusetts Institute of Technology, Cambridge, MA 02139, USA}
\affil[b]{Department of Biomedical Engineering, Boston University, Boston MA 02215, USA}
\affil[c]{Department of Pharmacology and Experimental Therapeutics, Boston University School of Medicine, Boston, MA 02118}
\affil[d]{Howard Hughes Medical Institute, Boston University, Boston, MA 02215, USA}
\leadauthor{Spill} 

\significancestatement{Oligonucleotide aptamers have increasing applications as a class of molecules that bind with high affinity and specificity to a target. Aptamers are typically selected from a large pool of random candidate nucleic acid libraries through competition for the target. Using a novel stochastic hybrid model, we are able to study the combined impact of important evolutionary success factors such as competition, randomness, and changes in the environment. While the environment may be tuned with experimental parameters such as target concentration, competition varies with differences in the initial distribution of aptamer-target binding affinities, and random events can eliminate even the ligands with the highest affinity.}

\authorcontributions{Author contributions: F.S., Z.B.W., D.D. and M.H.Z. designed research, F.S. developed the model, F.S. and D.D. implemented the model, F.S. and D.D. analyzed the data, Z.B.W, N.H., A.I.S. and D.D. contributed new reagents/analytic tools,  F.S., D.D. and M.H.Z. wrote the paper.}
\authordeclaration{The authors declare no conflict of interest.}
\correspondingauthor{\textsuperscript{1}To whom correspondence should be addressed. E-mail: ddesai@bu.edu (DD), zaman@bu.edu (MHZ)}


\begin{abstract}
The search for~\revised{high-affinity aptamers} for targets such as
proteins, \revised{small molecules, or cancer cells} remains a formidable
endeavor. Systematic Evolution of Ligands by EXponential Enrichment (SELEX) offers an iterative process to discover these aptamers through evolutionary selection of high-affinity candidates from a
highly diverse random pool. This randomness dictates an unknown
population distribution of fitness parameters, encoded by the binding
affinities, toward SELEX targets. Adding to this uncertainty,
repeating SELEX under identical conditions may lead to variable outcomes.
These uncertainties pose a challenge when tuning selection pressures
to isolate high-affinity ligands. Here, we present a novel stochastic
\revised{hybrid} model that describes the evolutionary selection of aptamers
in order to explore the impact of these unknowns. To our surprise, we
find that even single copies of high-affinity ligands in a pool of
billions can strongly influence population dynamics, yet their
survival is highly dependent on chance. We perform Monte Carlo
simulations to explore the impact of environmental parameters, such as the target concentration, on
selection efficiency in SELEX and identify new strategies to control
these uncertainties to ultimately improve the outcome and speed of
this time- and resource-intensive process.
\end{abstract}

\dates{This manuscript was compiled on \today}
\doi{\url{www.pnas.org/cgi/doi/10.1073/pnas.XXXXXXXXXX}}

\begin{document}

\verticaladjustment{-2pt}

\maketitle
\thispagestyle{firststyle}
\ifthenelse{\boolean{shortarticle}}{\ifthenelse{\boolean{singlecolumn}}{\abscontentformatted}{\abscontent}}{}


\revised{\dropcap{U}nderstanding and exploiting target-ligand binding are bedrocks of the biomedical sciences and support a host of applications ranging from diagnostics, therapeutics, and drug discovery to biosensing, imaging, and gene regulation. Antibodies and rational design provide a constructive playground to develop these applications, yet there generally remains a paucity of strong and specific binders for the innumerable viral, protein, and small molecule targets under investigation.

Aptamers offer an alternative to antibodies, yet in spite of their growth}~\cite{Shangguan08082006,ferguson2013real,keefe2010aptamers,bunka2006aptamers}, the discovery of~\revised{high-affinity} aptamers remains a challenge, especially for small molecule targets~\cite{mckeague2012challenges,blind2015aptamer}. Systematic Evolution of Ligands by EXponential Enrichment (SELEX)~\cite{tuerk1990systematic,ellington1990vitro}~\revised{is the premier framework for aptamer development and isolates high-affinity ligands from an initial library similar to how advantageous traits are enriched in a biological population through Darwinian selection. In a cyclic process, ligands are incubated with the target, and those that exhibit preferential binding are amplified and survive to the next round. Target molecules are typically immobilized on a substrate material to facilitate easy separation of target-bound and unbound ligands. Through numerous rounds of selection, an initial library can be reduced to a handful of high-affinity aptamers. Nucleic acids comprise the vast majority of libraries used in SELEX, where sequence regions are randomized to generate tremendous structural diversity. While this diversity underpins the evolutionary nature of SELEX, numerous works suggest that initial library design is a significant contributor to its overall success~\cite{luo2010computational}.}

\revised{While conceptually simple, the practical application of SELEX is plagued by uncertainty. Despite the impact of library design, the initial affinity distribution for any library toward a specific target remains} \textit{a priori} unknown. \revised{Target immobilization further complicates the procedure, particularly for small molecules.} In comparison to large molecular weight targets such as proteins~\cite{Cho12112013}, viruses~\cite{roh2011label}, and whole cells~\cite{farokhzad2006targeted,Sefah28012014}; the immobilization of small molecules eliminates ligand binding sites and is thus impractical. \revised{Newer approaches instead bind the library itself to a substrate material using non-covalent equilibrium binding, but this introduces the opportunity for competitive losses of high-affinity ligands that are initially present in extremely low numbers. Wash steps and other} experimental procedures may lead to further 	random losses, while non-specific selection of ligands can counter environmental pressures and stall selection. In short, these uncertainties may quickly compound to apply tremendous risk toward the guarantee of successful selection.

Mathematical modeling therefore has great potential to help understand the uncertainties of aptamer selection and devise strategies to optimize environmental parameters and improve selection outcomes. Previous models have explored SELEX for protein targets, considering parameters such as target concentration~\cite{irvine1991selexion,levine2007mathematical,wang2012influence}, separation efficiency of target-bound and unbound ligand~\cite{chen2007subtractive}, nonspecific binding of DNA to target~\cite{cherney2013theoretical}, and negative selection steps~\cite{seo2014computational}. \revised{These studies predict that, in spite of its experimental complexity, the evolutionary nature of SELEX guarantees selection of the highest affinity ligand from the initial library. However, these works focus primarily} on the use of deterministic equilibrium equations~\cite{irvine1991selexion}, whereas the presence of ligands in low copy numbers and the role of other experimental uncertainties \revised{suggest} the use of more fundamental stochastic models rather than deterministic approximations. \revised{Mathematically,} the chemical master equation provides a framework to \revised{test this hypothesis} and generalize the above-mentioned deterministic models to include intrinsic stochasticity~\cite{van1992stochastic}. While this \revised{approach} could be applied toward a purely stochastic model for SELEX, the result cannot currently be solved analytically or simulated by conventional techniques such as the Gillespie algorithm~\cite{gillespie1976general}, due to the large number of molecules present. These limitations are common for many stochastic multiscale problems in biology, chemistry and physics; and the development of novel analytic approximations or numerical techniques to address this problem is an important ongoing research topic~\cite{Gillespie2014Perspective}.

Using these ideas as our foundation, we introduce a new \revised{hybrid} model for aptamer selection that builds on the chemical master equation to introduce stochastic uncertainty in SELEX modeling. Here, ligands are separated into two categories of high and low copy number. In the former case, the master equation is simplified toward a deterministic equilibrium system, whereas in the latter it can be approximately solved analytically. Unlike previous efforts to incorporate stochasticity into aptamer modeling~\cite{sun1996mathematical,chen2007complex}, our framework allows us to simultaneously investigate the impact of low copy number ligands \revised{and their competitive binding to target molecules and immobilization substrates among the presence of high copy number ligands. Most importantly, this approach can capture total loss of individual ligands, which can strongly contribute to protocol outcome. Such events have not previously been investigated and cannot be captured by other approximations of the master equation such as the Langevin approximation, which rely on the presence of sufficiently high numbers of molecules and thereby diminish the possibility of extinction events \cite{gillespie2000chemical}.}

\revised{Using this framework,} we investigate unexplored sources of uncertainty in SELEX, \revised{beginning with a systematic analysis of the role the initial library affinity distribution plays in selection}. We further challenge the assumption that this distribution is continuous at its tails and evaluate the impact of adding noise at these extremes. We find that introducing as few as $20$ additional ligands outside the bulk distribution of $10^{15}$ molecules can strongly affect the outcome of selection. In light of these results, we revisit the topic of optimizing target concentration as discussed in previous works~\cite{irvine1991selexion,levine2007mathematical,wang2012influence}, and show that the assumed initial $K_D$ distribution strongly influences~\revised{protocol optimizations}. We also provide additional insights regarding non-covalent ligand immobilization to support more recent efforts to develop robust protocols for small molecule SELEX~\cite{stoltenburg2012capture,wooakim2012immobilization,seopakwon2014multiple}. \revised{Integrating these ideas, we show that simultaneously lowering the target concentration and the substrate binding dissociation constant over the SELEX cycles can lead to improved selection outcomes for a wide range of initial conditions.}

\section*{Computational Model of Selection Dynamics}

\begin{figure}
\includegraphics[width=0.96\linewidth]{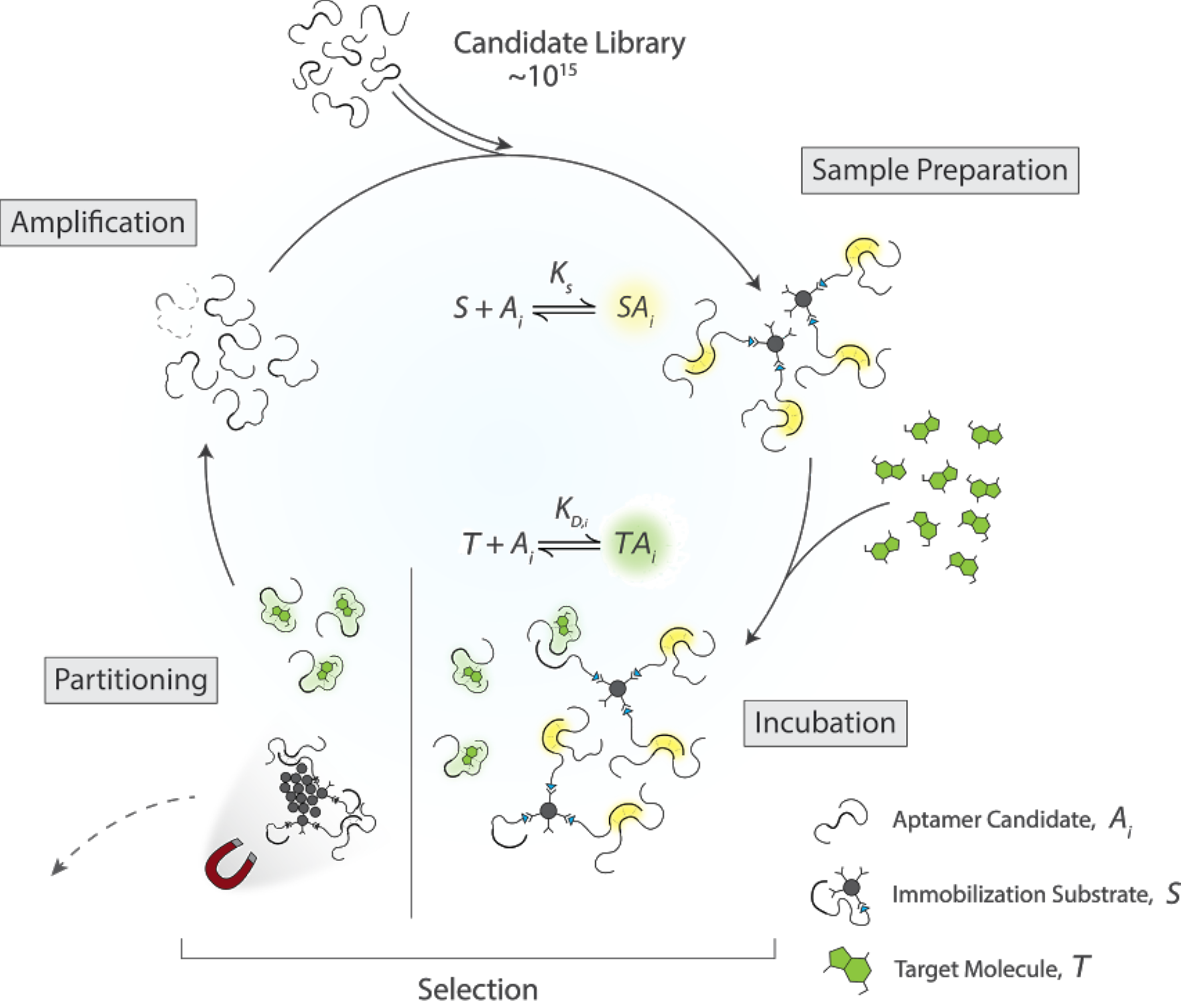}
\vspace{-8pt}
\caption{\label{fig:capture_selex}
A sample candidate library of ligands $A_i$ is prepared by letting the ligands bind to a substrate $S$. Then, the target is added, leading to competitive binding between the different apatamers for substrate and target molecules $T$. The ligands still bound to the substrate are then separated from those which are either bound to a target, or have randomly unbound from the substrate. The latter two are subsequently amplified and taken into the next cycle.
}
\end{figure}

\revised{The original SELEX protocol~\cite{tuerk1990systematic,ellington1990vitro} serves as the basis for our model}, with additional modifications to accommodate small molecule targets as described in \cite{stoltenburg2012capture}. While this marks the first model that specifically considers small molecule targets, the main ideas and conclusions derived from this work remain applicable for other targets and selection schemes. The main steps of our approach are summarized in Fig. \ref{fig:capture_selex}. We begin with a library of $\Anum_i^{tot}$ ligands of type $i$, where $i=\{1,\dots,M^{A}\}$ and $M^{A}$ is the total number of unique ligands. The ligands are then \revised{non-covalently immobilized using $\Snum$ substrate molecules, where} $K_S$ is the ligand-substrate dissociation constant. \revised{These complexes are then} subjected to wash steps to remove unbound ligands, from which $\Anum_i^{I}$ ligands of type $i$ survive. \revised{Surviving} ligands are then incubated with $\Tnum$ target molecules, where a ligand of type $i$ binds to the target with a dissociation constant $K_{D,i}$. Ligands that are bound to a target or have unbound from the substrate are partitioned from those that remain bound to the substrate. Finally, the partitioned ligands are amplified \textit{via} PCR, modeled as a constant factor increase of $\alpha_{PCR}$, and used to begin the next cycle. The proceeding sections highlight the notable details of our \revised{hybrid} approach, while a more thorough description and derivation of the model can be found in the supporting information. Throughout these sections, quantities that refer to an absolute number of molecules are denoted with a tilde, while those without represent concentrations.

\subsection*{Deterministic Model of Ligand Binding}\label{sec:specificDeterministicBinding}

Earlier works \revised{use equilibrium conditions} to characterize ligand-target interactions \revised{during} selection~\cite{irvine1991selexion,levine2007mathematical,chen2007subtractive,wang2012influence}\revised{, focusing on changes in bulk properties, such as the mean dissociation constant, to study the enrichment of a single best candidate. We instead monitor the full ligand affinity distribution} in an effort to better understand how parameters such as the initial standard deviation also impact selection dynamics. Since modeling each of the $M^A\approx10^{15}$ unique ligands is computationally intractable, we discretize the initial distribution of $M^A$ unique ligands into $M^B$ bins, each containing $\Anum_i$ ligands of dissociation constant $K_{D,i}$, where $i=\{1,\dots,M^B\}$. We choose $M^B$ to be large enough that the results do not depend on the binning, and small enough to optimize simulation performance. We further build on this analysis by introducing additional equilibrium conditions for non-specific ligand-substrate interactions represented by a dissociation constant $K_S$. \revised{In \cite{stoltenburg2012capture}, substrate-ligand binding is accomplished through DNA base pairing using a fixed sequence, and is thus constant. Altering the length of this fixed sequence is a means to tune $K_S$. Moreover, different immobilization techniques, such as the use of graphene oxide~\cite{wooakim2012immobilization,seopakwon2014multiple}, will lead to variations of $K_S$ within a given pool, but we do not consider such cases here and consider $K_S$ to be constant throughout a single cycle of SELEX. Combining ligand-target and ligand-substrate binding,} the full system of steady-state equilibrium binding conditions can be described by the set of equations:

\begin{align}\label{eq:deterministicSelection}
\begin{split}
[S A_i] &= \frac{1}{K_{S}}(A_i^{I}-[S A_i]-[TA_i])S^{free},\quad i = 1,\dots ,M^B, \\
[T A_i] &= \frac{1}{K_{D,i}}(A_i^{I}-[S A_i]-[TA_i])T^{free},\quad i = 1,\dots ,M^B,\\
S^{tot} &= \sum_{i=1}^{M^B}[S A_i] + S^{free},\quad T^{tot} = \sum_{i=1}^{M^B}[T A_i] + T^{free} 
\end{split}
\end{align}
Here, $[S A_i]$ and $[T A_i]$ denote the concentration of ligand-substrate and ligand-target complexes, \revised{representing $2M^{B}$ independent variables that are solved; the quantities $T^{tot}, T^{free}$ and $S^{tot},S^{free}$ denote the concentrations of total and free target and substrate, respectively.} From these results, we determine the concentration of ligands which survive selection, denoted by $A_i^{S,D}$, and are amplified by PCR for the next cycle. The superscripts denote that this number is obtained after selection and using the deterministic model defined by \eqref{eq:deterministicSelection}. This concentration is simply the sum of free and target-bound ligands, and is hence given by 
\begin{equation}\label{eq:concentrationSelectedAptamers}
A_i^{S,D} = [T A_i] + A_i^{free} = A_i^{I}-[S A_i].
\end{equation}
%

\subsection*{Stochastic Model of Ligand Selection}\label{sec:stochasticModel}

Chemical reactions are fundamentally stochastic in nature, with forward and backward reactions occurring constantly. While powerful and simple, \revised{\eqref{eq:deterministicSelection} is based on real-valued concentrations which require sufficiently high molecular copy numbers to make discreteness and random fluctuations negligible.} This is \revised{challenged} at the tails of the $K_D$ distribution, where appropriate binning results in few ligands per bin. To address this, a \revised{hybrid} approach is used where additional stochastic analysis is applied when \eqref{eq:deterministicSelection} predicts $\Anum_i^{S,D}$ to be below a threshold $\Theta$. To distinguish these quantities for stochastic analysis, we denote them as $\Anum_\psi^{S,D}$, where $\psi$ represents the subset of indices $i$ that satisfy the condition $\Anum_i^{S,D}<\Theta$. \revised{Results exploring the choice for $\Theta$ are provided in SI Fig. S6.} We then calculate the probability for selecting $\Anum_\psi^{S,S}$ ligands, $p\left(\Anum_\psi^{S,S}\right)$; the superscripts denotes that the number is obtained after selection and using the stochastic model. As described in the supporting information, we find that by starting with the chemical master equation, $p\left(\Anum_\psi^{S,S}\right)$ is well-approximated by a binomial distribution:
\begin{align}\label{eq:aptamerBinomialSelection}
\begin{split}
p\left(\Anum_\psi^{S,S}\right) &= { \Anum_\psi^{tot} \choose \Anum_\psi^{S,S}} p_\psi^{\Anum_\psi^{S,S}} (1-p_\psi)^{\Anum_\psi^{tot} - \Anum_\psi^{S,S}}, \\ 
\mbox{for } \Anum_\psi^{S,S} &= 0,\dots,\Anum_\psi^{tot}
\end{split}
\end{align}
Here, the quantity $p_\psi$ represents the probability that a single ligand is selected out of $\Anum_\psi^{tot}$ ligands of type $\psi$. To provide the most accurate description, we account for stochastic contributions from both the immobilization and incubation steps. The contribution from immobilization is approximately the same for all candidates, and is given by $\frac{\Anum^{I}}{\Anum^{tot}}$, the fraction of remaining immobilized ligands after wash steps over those present before immobilization, where $\Anum^{I}=\Sigma_{i=1}^{M^B}{\Anum_i^{I}}$ and $\Anum^{tot}=\Sigma_{i=1}^{M^B}{\Anum_i^{tot}}$. The contribution from incubation is calculated as the fraction of predicted ligands, $\Anum_\psi^{S,D}$, out of an initial number of $\Anum_\psi^{I}$. Using these contributions, the total probability that a ligand in bin $\psi$ survives is given by:

\begin{align}\label{eq:probabilitySelectSingleAptamer}
p_\psi = \frac{\Anum^I A_\psi^{S,D}}{\Anum^{tot} A_\psi^{I}}
\end{align}
Finally, \eqref{eq:aptamerBinomialSelection} requires $\Anum_\psi^{tot}$ to be integer-valued, as it denotes a number of molecules. However, the deterministic equations yield real-valued concentrations that must be renormalized to an integer. We separate $\Anum_\psi^{tot}$ into its integer and fractional parts, $\Anum_\psi^{tot} = \Anum_{\psi,\mathbb{N}}^{tot} + \Anum_{\psi,f}^{tot}$, and then interpret $0\leq\Anum_{\psi,f}^{tot}<1$ as the probability to have an extra molecule present. We then draw a uniformly distributed random number $0\leq r \leq 1$, and set $\Anum_{\psi}^{tot} =  \Anum_{\psi,\mathbb{N}}^{tot} + 1$ if $r <\Anum_{\psi,f}^{tot}$, and $\Anum_{\psi}^{tot} =  \Anum_{\psi,\mathbb{N}}^{tot}$ otherwise. Following this renormalization, we finally draw a random variate distributed according to \eqref{eq:aptamerBinomialSelection} to simulate the set of ligands $\Anum_\psi^{S,S}$ that remain after both immobilization and selection.

\section*{Results and Discussion}\label{sec:results}

Utilizing a \revised{hybrid} computational approach, our model provides a generalized framework that can be used to analyze both deterministic and stochastic effects in SELEX. We use the model to deconstruct two main forms of uncertainties in aptamer selection. The first is parameter uncertainty, including the unknown initial $K_D$ distribution \revised{as well as the experimentally tunable quantities $K_S$ and $T^{tot}$. These are analyzed} using a parameter study that observes the impact of these factors on SELEX dynamics. The second is stochastic uncertainty associated with low copy number binding phenomena. As this form of uncertainty is random in nature, we employ Monte Carlo simulations to observe the variability in outcomes between repeated SELEX procedures and extract conclusions which are robust with respect to stochastic fluctuations. Unless mentioned otherwise, the parameters from Table S1 are used in all simulations.

\subsection*{Effect of $K_{D}$ Distribution on Selection Efficiency}\label{sec:ImpactKdDistribution}

\begin{figure}
\centering
\includegraphics[width=0.98\linewidth]{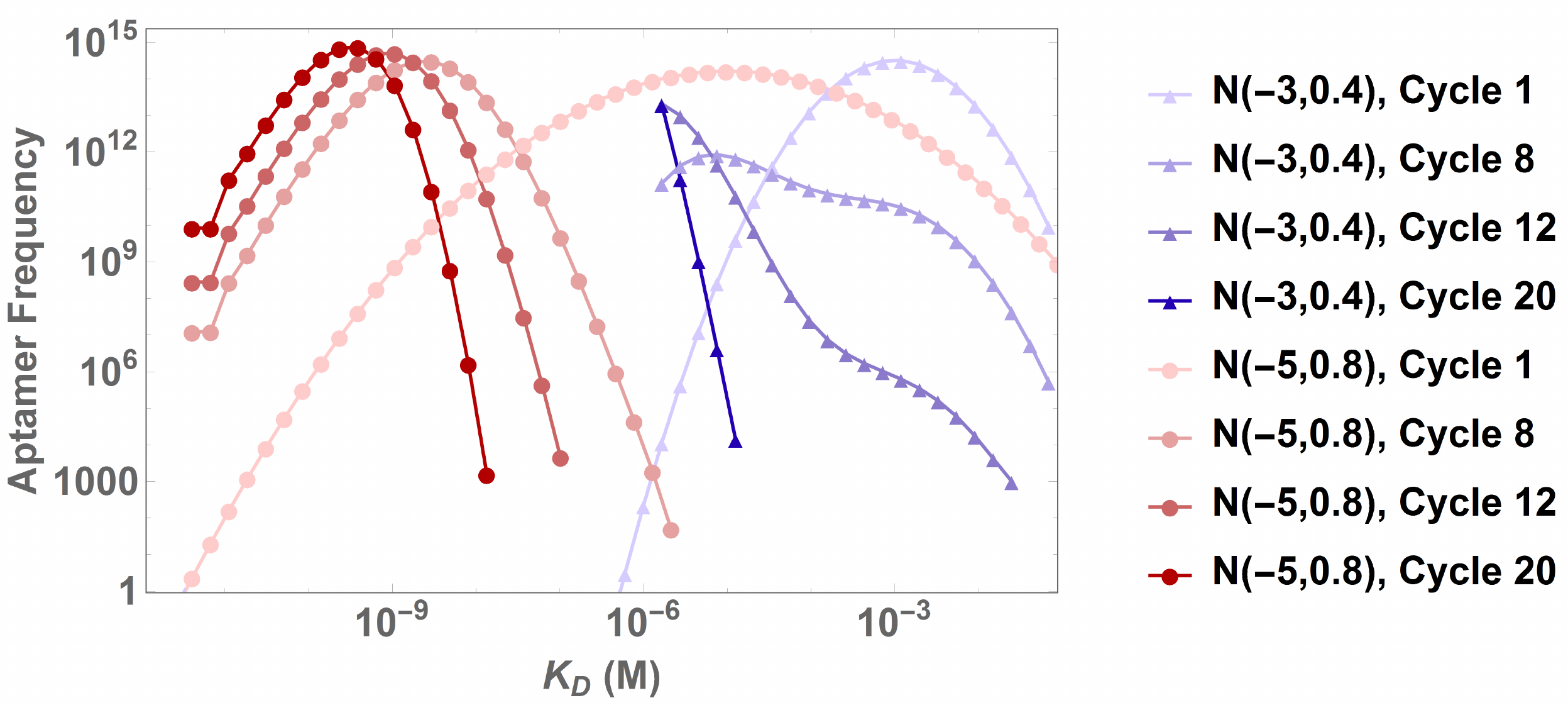}
\vspace{-10pt}
\caption{
Initial Distribution affects SELEX dynamics. We plot the distribution of ligand binding affinities with increasing SELEX cycles for the same experimental parameters and two different assumed Gaussian distributions at cycle $1$, $N(-3,0.4)$ (blue triangles) and $N(-5,0.8)$ (red dots). The dynamics of the two cases are totally different. For $N(-5,0.8)$, the distribution shifts to the left and becomes considerably narrower, and for $N(-3,0.4)$, the distribution additionally skews to the left, such that from cycle $12$ on the highest-affinity binders have outcompeted the rest of the distribution.
}
\label{fig:DistributionsNoNoise}
\end{figure}

Gaussian distributions describing the initial ligand pool dominate SELEX models in literature~\cite{wang2012influence}, yet we are not aware of any prior systematic approach to study the impact of various distributions on the outcome of SELEX. While strong justifications have been made for the assumption of a log-normal Gaussian description~\cite{vant1998mathematics}, we explore various Gaussian as well as non-Gaussian distributions and their impact on selection. Our convention for log-normal $K_{D}$ distributions is such that a Gaussian $N(\mu,\sigma)$ with mean $\mu$ and standard deviation $\sigma$ in log-space translates \revised{to a} mean of $10^\mu$ in $K_{D}$ space; we do not shift the mean by $\frac{1}{2}\sigma^2$ as is customary in Ito calculus. Fig. \ref{fig:DistributionsNoNoise} highlights the dramatic difference observed for just two different assumed distributions\revised{, and demonstrates the significant role the initial $K_D$ distribution plays in SELEX. This point is further accentuated by the fact that different selection targets may significantly alter the initial $K_D$ distribution for any given library.} SI Fig. S1 confirms that for a variety of other distributions, including non-Gaussians, distribution shape has a dramatic impact on selection dynamics.

In addition to shape, we also explore the assumption that the $K_D$ distribution is continuous everywhere. While this assumption is credible near the distribution mean where the frequency of molecules is sufficiently high, we expect it to fail at the extreme tail where stochastic effects dominate and highly specific sequences can create gaps in the affinity distribution. Indeed, it is well-known that even single base-pair changes in DNA can dramatically impact binding \cite{katilius2007exploring}. Ligands in this regime are highly prized, but may also be at highest risk to be lost to stochastic effects due to low copy numbers.

\begin{figure}
\centering
\includegraphics[width=0.98\linewidth]{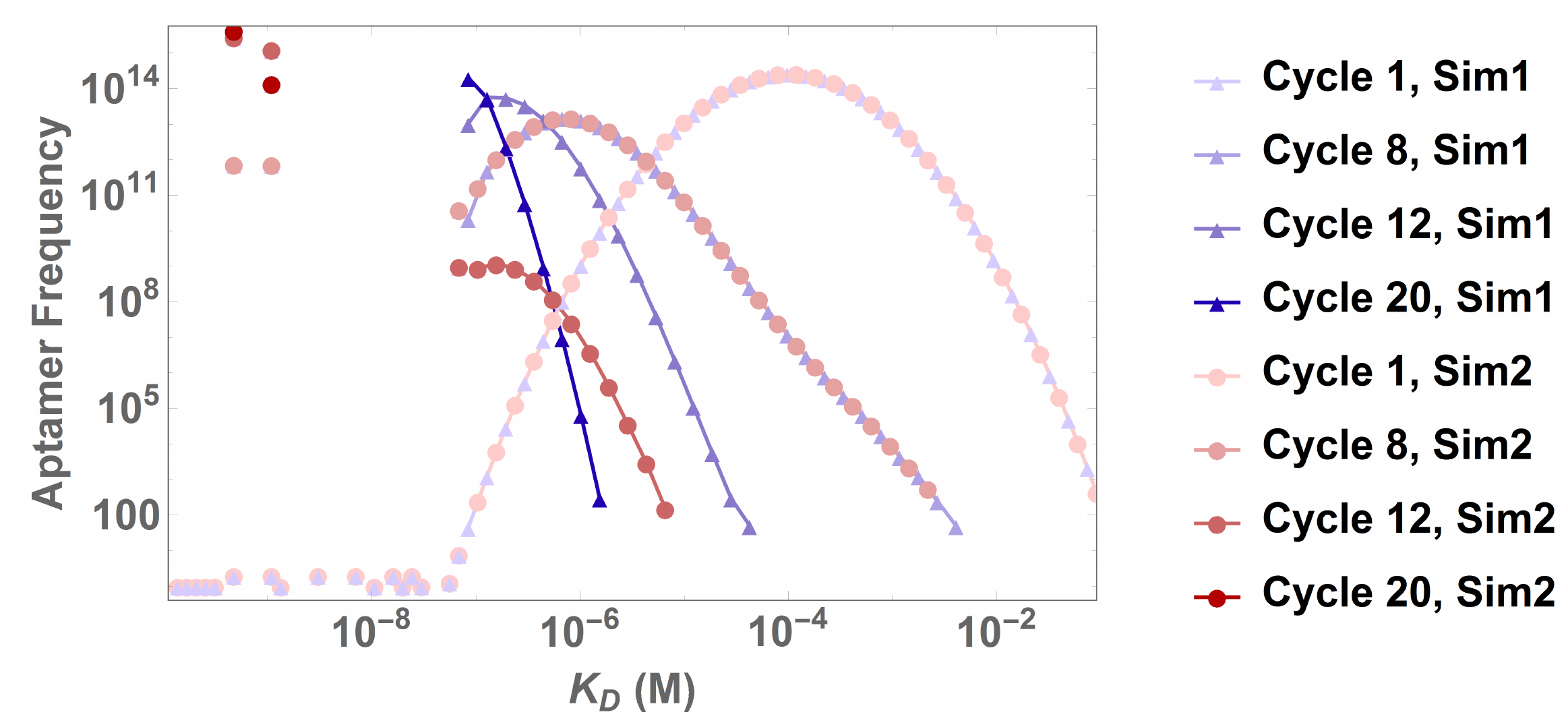}
\vspace{-10pt}
\caption{\label{fig:DistributionsWithNoise}
Noise affects SELEX dynamics. We fix the experimental parameters, and the same initial Gaussian distribution $N(-4,0.4)$ with the same added noise of only $20$ additional ligands initially present between $K_{D}=10^{-10}M$ and $5\times 10^{-8}M$. Two different Monte Carlo simulations show dynamics of selection under random loss of the $20$ strongest binders, (blue triangles), versus dynamics when only two of those strong binders with affinities between $10^{-10}$ and $10^{-9}M$ are selected, (red dots). In the latter cases, these two high-affinity binders completely dominate the distribution from Cycle $12$ on and outcompete the remaining ligands with low affinities ($K_{D}>10^{-7}M$).
}
\end{figure}

We investigate this risk by using an initial $N(-4,0.4)$ distribution and adding a fixed noise component that is randomly sampled from a uniform distribution in log-space. Fig. \ref{fig:DistributionsWithNoise} and Movie S1 show a comparison of two Monte Carlo simulations where there are only $20$ ligands present in the range of $K_{D}<10^{-7}M$, i.e. where the continuous Gaussian distribution is effectively zero. \revised{We find that random} binding effects can lead to total loss of those $20$ ligands, \revised{resulting in a very different evolution of the $K_D$ distribution from cycle 12 onward in comparison to the case where only $2$ of those ligands survive}. SI Fig. S2 shows a distribution of the mean ligand $K_D$ at cycle $20$ obtained from $250$ Monte Carlo simulations, confirming this enormous variability in outcomes, \revised {where the mean $K_D$ value spans three orders of magnitude}.

These results demonstrate the tremendous sensitivity of selection dynamics to both distribution shape and noise. They illustrate that selection pressures are parameterized not only by extrinsic environmental conditions given by the experimental setup, such as \revised{the tunable quantities $K_S$ and $T^{tot}$}, but just as importantly by \revised{inherently uncertain} intrinsic population parameters that govern relative competition between ligands of varying affinities.

\subsection*{Revisiting Target Concentration}

Optimization of the target concentration, $T^{tot}$, has long stood as a critical step in adjusting selection pressure based on experimental parameters~\cite{irvine1991selexion,levine2007mathematical,wang2012influence}. However, the results from the previous section now suggest that in addition to these experimental factors, the intrinsic affinity distribution of the initial ligand pool may have a significant influence on the impact $T^{tot}$ exerts on the overall selection pressure. In light of this, we revisit the topic to study this impact by varying both $T^{tot}$ and the initial distribution. 
\begin{figure}
\centering
\includegraphics[width=0.96\linewidth]{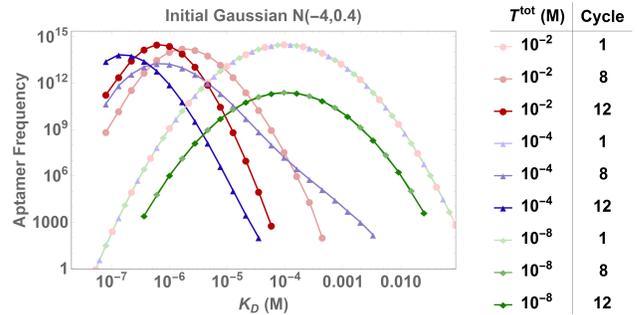}
\vspace{-10pt}
\caption{\label{fig:InitialDrugConcDependenceTimeEvolution}
Impact of target concentration on SELEX dynamics. Evolution of $K_{D}$ distribution for three different values of the target concentrations is shown. Under a high target concentration of $T^{tot}=10^{-2}M$, the distribution shifts to the left and narrows, but does not skew towards high-affinity ligands. Additional skewing is achieved by reducing to $T^{tot}=10^{-4}M$, which increases selection pressure by intensifying ligand competition. However, further reduction to $T^{tot}=10^{-8}M$ has the opposite affect and actually halts selection. In this case, the target concentration is so low that non-specific ligand-substrate equilibria dominate selection dynamics and nullifies the selection pressure.
}
\end{figure}
\begin{figure}
\centering
 \includegraphics[width=0.96\linewidth]{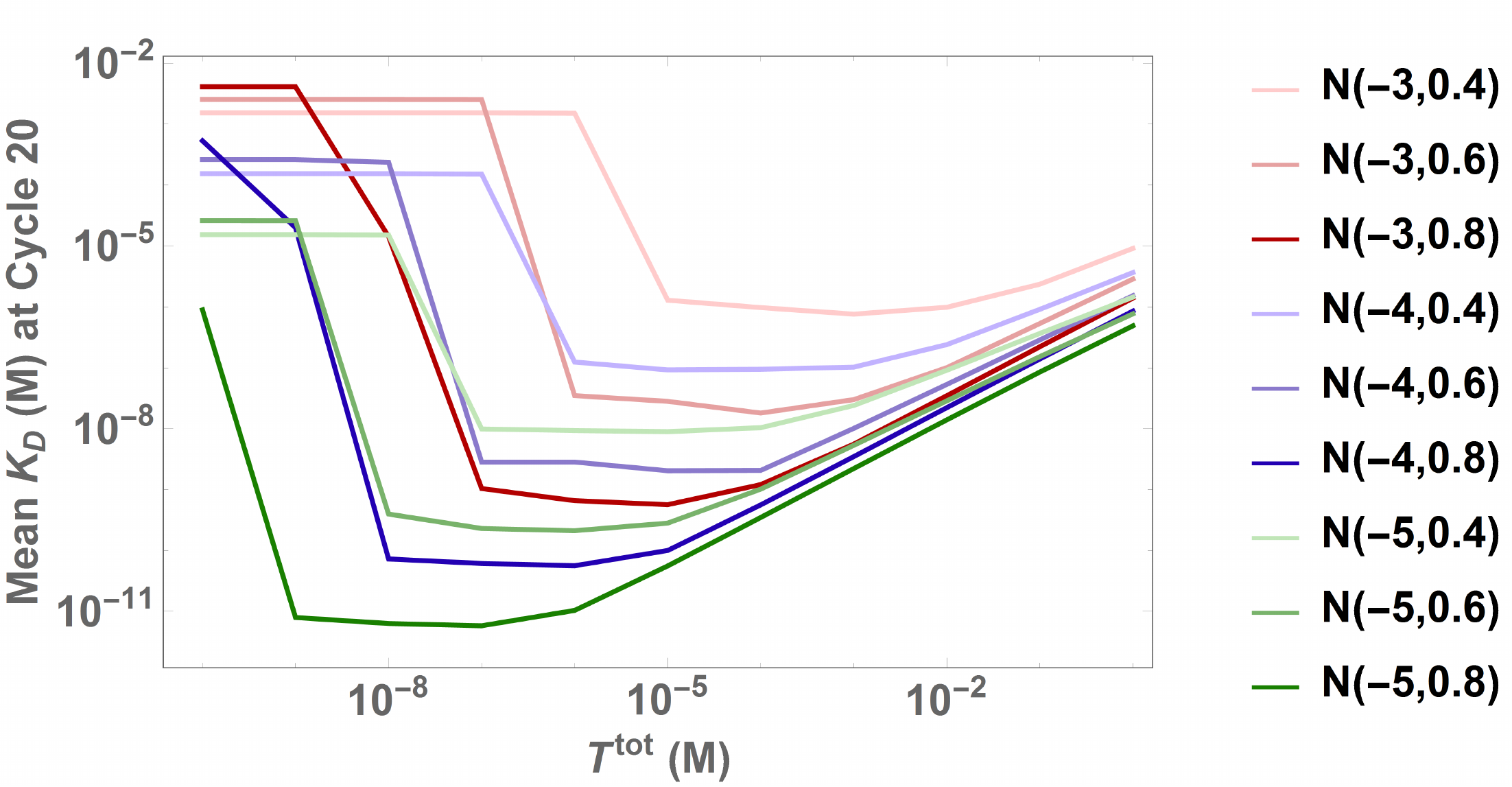}
\vspace{-10pt}
\caption{\label{fig:InitialDrugConcDependence}
Optimal target concentrations strongly depend on assumed initial $K_{D}$ distribution. Mean $K_{D}$ \revised{as a measure of pool binding strength} for SELEX pool at cycle $20$ using different constant target concentration. Depending on the initial distribution of ligands, we find vastly different optimal target concentrations, i.e. concentrations with lower mean $K_D$.
}
\end{figure}
\revised{
Fig. \ref{fig:InitialDrugConcDependenceTimeEvolution} and Movie S2 first show the dramatic impact of drug concentration on selection dynamics. The results indicate that $T^{tot}=10^{-4}M$ (blue) provides optimal selection out of the three investigated drug concentrations that use the initial Gaussian distribution $N(-4,0.4)$. To investigate the impact of $T^{tot}$ more systematically, Fig. \ref{fig:InitialDrugConcDependence} shows the mean $K_{D}$ value of ligands selected after $20$ cycles as a function of $T^{tot}$ for nine different initial distributions. Note that as the mean $K_D$ decreases, the average binding strength of the pool increases. Fig. \ref{fig:InitialDrugConcDependence} confirms that intermediate values of $T^{tot}$ yields optimal selection. SI Figs. S7 (a)-(c) further show that adding noise to the initial distributions introduces additional variability, but provides similar qualitative results. Interestingly, we find that different initial distributions can have very different optimal $T^{tot}$, stressing the importance of devising a strategy to mitigate this impact and thereby control the inherent uncertainty associated with the initial $K_D$ distribution.}

\subsection*{$K_{S}$ Dependence and Non-specific Selection}

Our \revised{hybrid} model has allowed us to explore the impact of the unknown initial $K_D$ distribution and the target concentration $T^{tot}$, which are both present in all SELEX protocols. However, our model additionally introduces a ligand-substrate interaction that has never before been studied  and offers a unique opportunity to apply it toward more recent selection schemes aimed at small molecule aptamer development~\cite{stoltenburg2012capture,wooakim2012immobilization,seopakwon2014multiple}. We therefore extend our analysis to study uncertainties that govern an optimum $K_S$, and observe how changes in $K_S$ impact selection dynamics for different $K_D$ distributions.

\begin{figure}
\centering
\includegraphics[width=0.96\linewidth]{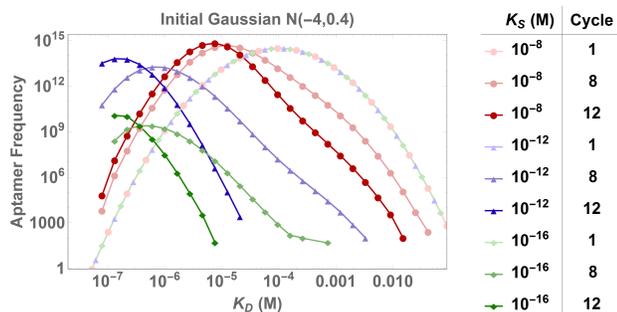}
\vspace{-10pt}
\caption{\label{fig:InitialKOligoDependenceTimeDependence}
Impact of $K_S$ on SELEX dynamics. Evolution of $K_{D}$ distribution for three different values of $K_{S}$. Similar to target concentration, we find an optimal outcome in the middle range ($K_S=10^{-12}M$, blue), but the outcome for low $K_S$ is not as adverse as for low $T^{tot}$, since the distribution still shifts towards low $K_D$ with increasing cycles.
}
\end{figure}
\begin{figure}
\centering
\includegraphics[width=0.96\linewidth]{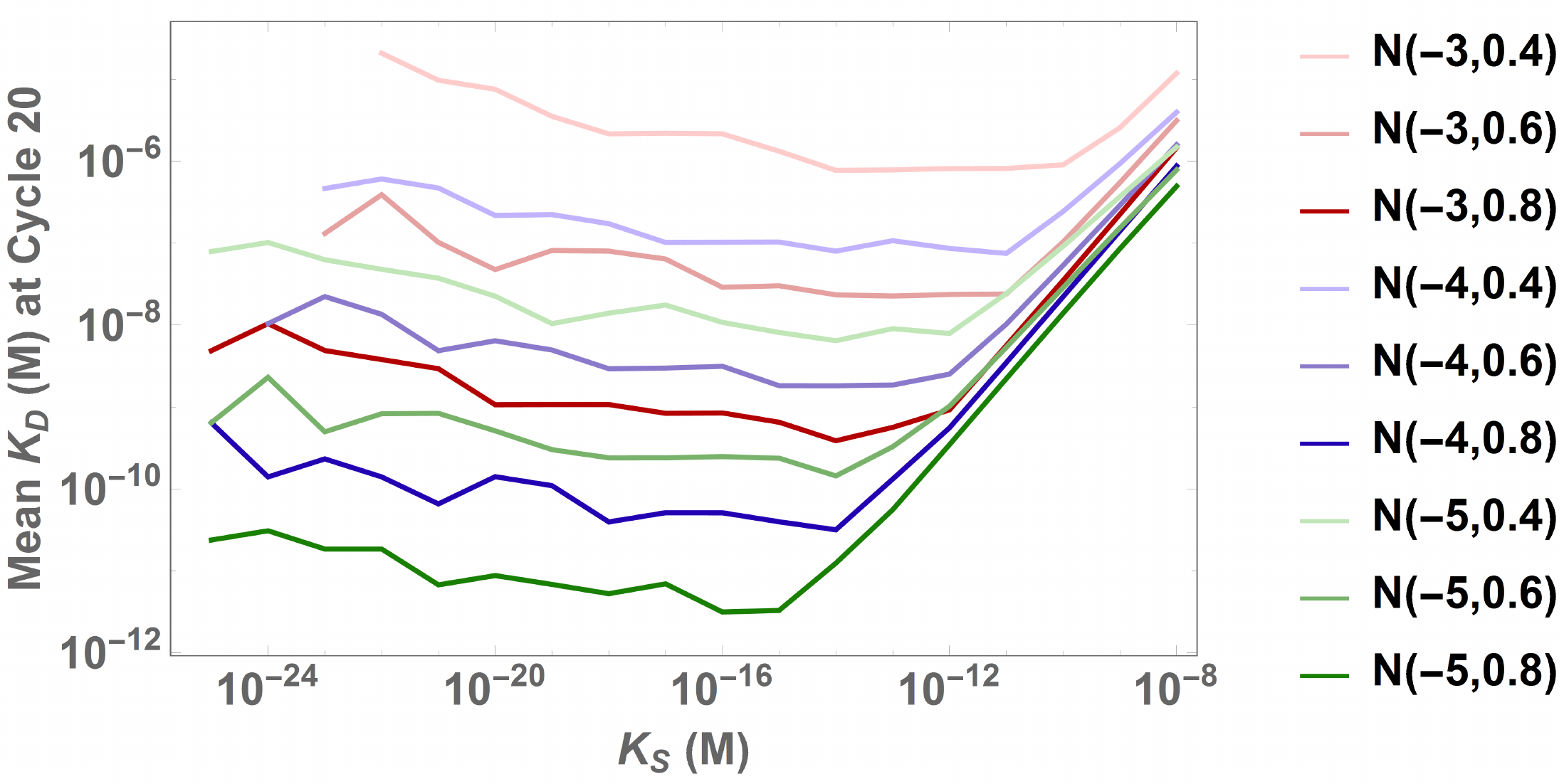}
\vspace{-10pt}
\caption{\label{fig:InitialKOligoDependence}
Optimal $K_S$ depends on initial distribution. Plot of mean $K_{D}$ for SELEX pool at cycle $20$ using different values of $K_{S}$. Reducing $K_S$ from its optimal value does not increase the mean $K_D$ as strongly as a reduction of the target concentration from its optimum, as shown in Fig. \ref{fig:InitialDrugConcDependence}.
}
\end{figure}
\revised{
Fig. \ref{fig:InitialKOligoDependenceTimeDependence} and Movie S3 show the evolution of a single initial $K_{D}$ distribution for three different values of $K_S$, showing an optimal outcome for $K_{S}=10^{-12}M$ (blue). Noting these dynamics, we next vary $K_S$ systematically and observe the mean $K_{D}$ value of ligands present at cycle $20$ for nine different initial $K_D$ distributions (Figs. \ref{fig:InitialKOligoDependence}, S7(d)-(f)). Similar to target concentration, we find an optimum in the intermediate ranges of $K_{S}$ and a clear dependence on the initial distribution. However, contrary to target concentration, the mean $K_D$ for smaller $K_{S}$ is relatively insensitive. Thus, these results suggest that a lower value of $K_{S}=10^{-16}M$ would provide similar results across a multitude of initial distributions.

As it pertains to small molecule selection schemes, these results provide useful insights into the impact that substrate binding affinity has on selection efficiency, and may offer some guidance in the appropriate selection of a substrate material. The results also provide general insights into the impact of partitioning efficiency and non-specific binding on selection across various initial distributions and suggest that a given partitioning efficiency or fraction of non-specific selection can impact different initial distributions in vastly different ways. 
}

\subsection*{Improving Selection Efficiency}
We have shown that the initial $K_{D}$ distribution has a tremendous impact on selection efficiency and plays a significant role in modulating the impact of experimental parameters such as $T^{tot}$ and $K_{S}$. These results highlight that while established protocols are expected to perform well for some distributions, they may perform moderately for others. To address this variability in outcomes, we finally explore strategies to mitigate these impacts using only the experimental parameters $T^{tot}$ and $K_S$. As a metric for our analysis, we introduce the quantity $\phi(c)$, which describes the fraction of ligands with $K_D < 10^{-10}M$ at cycle $c=\{1,\dots,C\}$. Using this quantity, we further introduce two measures of efficiency: success probability $\Phi=\phi(C)$ and success speed $S_C$ defined as the cycle $c$ at which $\phi(c)=0.5\phi(C)$.

We have seen that $K_{S}$ and $T^{tot}$ play distinct roles in the evolutionary dynamics of the $K_D$ distribution. \revised{However, both parameters exhibit regimes of optimal selection that depend heavily on the initial distribution mean and width. Figs. \ref{fig:InitialDrugConcDependence} and \ref{fig:InitialKOligoDependence} show that high values for $T^{tot}$ and $K_{S}$ have a similar impact across all distributions, and suggest a conservative approach of beginning at these high values for the initial cycles. This reduces the risk of eliminating high-affinity, low copy number ligands early on. As these high-affinity ligands are amplified in subsequent rounds,} $T^{tot}$ and $K_{S}$ can be lowered to rapidly eliminate the remaining low-affinity ligands (see SI Figs. S3, S4). \revised{While ideas to lower the target concentrations have been discussed previously \cite{stoltenburg2012capture}, our results indicate that other parameters such as $K_S$ can be tuned simultaneously to improve outcome across a multitude of initial distributions and stochastic conditions.}
\begin{figure}
\centering
\includegraphics[width=0.96\linewidth]{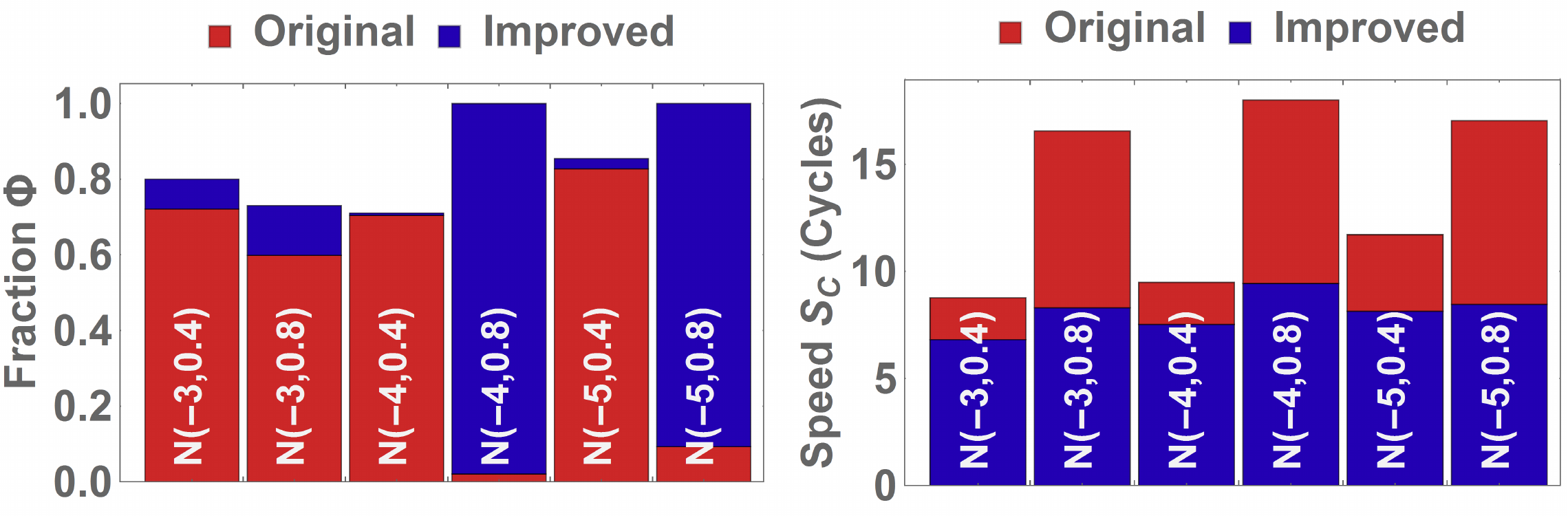}
\vspace{-5pt}
\caption{\label{fig:ImprovedProtocolPlateauAndSpeed}
Plots comparing the fraction of high-affinity ligands $\Phi$ and speed $S_C$ of SELEX for six different $K_{D}$ distributions. The values are obtained from averaging $50$ Monte Carlo simulations. We observe that the protocol with decreasing $T^{tot}$ and $K_{S}$ over the rounds will lead to a higher fraction of strong binders (here, with $K_{D} < 10^{-10}M$), and will reach this fraction faster, than the protocol where $T^{tot}$ and $K_{S}$ are kept constant.
}
\end{figure}
Fig. \ref{fig:ImprovedProtocolPlateauAndSpeed} shows $\Phi$ and $S_C$ obtained from $50$ Monte Carlo simulations of an \revised{improved} protocol where both $T^{tot}$ and $K_{S}$ are decreased over the cycles as described in Table S2. These results are compared to the original protocol with constant values $T^{tot}=10^{-4}M$ and $K_{S}=10^{-12}M$~\cite{stoltenburg2012capture}; SI Fig. S5 shows $\phi(c)$ including the standard deviations. Using six different initial Gaussian distributions with noise added similar to Fig. \ref{fig:DistributionsWithNoise}, we observe that the \revised{improved} protocol with decreasing $T^{tot}$ and $K_{S}$ \revised{is faster and leads to a higher fraction of high affinity binders than the original protocol. As an alternative metric of protocol performance, SI Fig. S8 shows the evolution of mean $K_D$ across the cycles, and also introduces two alternative protocols where $T^{tot}$ or $K_S$ are decreased faster than in the improved protocol. The results indicate that while faster decreases can further improve performance for some distributions, they may also lead to adverse outcome for others.}

\section*{Conclusions and Outlook}\label{sec:Conclusions}
\revised{
Deterministic models for SELEX have shed tremendous insight on the challenges faced in aptamer selection, but have been unable to capture its inherently uncertain nature. Here, we have presented a \revised{hybrid} model that captures stochastic binding and furthermore incorporates non-covalent ligand-substrate immobilization. Using this framework, we have investigated previously unexplored questions including the role of the initial library $K_D$ distribution, impact of distribution noise, and the effect of these factors on the optimization of experimental parameters such as the total target concentration $T^{tot}$ and the substrate dissociation constant $K_S$.

The results of our modeling draw striking parallels to outcomes in evolutionary biology,} where environmental parameters define a fitness landscape and competition can change this landscape to influence survival and reproduction \cite{nowak2004evolutionary}. Within SELEX, ligands compete for target molecules to ensure survival into the next cycle, whereas substrate binding traps the ligands and leads to their removal. Reduction of target concentration can increase competition, but when few target molecules are present, even high-affinity binders are unlikely to find a target. Similar to competition in limited resources scenarios, we find that the chance of survival for even the highest affinity ligand strongly depends on the strengths of the other ligands present in the population. \revised{Our surprising finding that a handful of high-affinity ligands can outcompete a pool of $10^{15}$ ligands is also seen in evolutionary biology, where highly advantageous traits can quickly spread in a population, given the right conditions. The model enables one to identify the parameters impacting selection, and can thus be used to improve selection efficiency.} A further important component of evolution in biological systems is mutations. Mutations in SELEX can also appear during PCR amplification, but usually lead to reduced affinities of the strongest aptamers \cite{katilius2007exploring}, so we ignored them in our current approach. However, for some SELEX protocols, mutations can be beneficial to expand the experimental sampling space \cite{hoinka2015large}, and it may be interesting to extend our model to those protocols.

In summary, our novel model provides a better understanding of the impact of the uncertainties in SELEX, and how experimental parameters can be tuned to improve outcome and speed of this expensive and time-consuming protocol. We have demonstrated how optimization of the parameters can enhance selection efficiency of one protocol dramatically, and we envisage that simple adaptations of our model can be used to improve the many other established protocols, as well as guide the design of novel protocols, which aim to limit the impact of uncertainties in selection methods.
%


\acknow{We acknowledge the support of the NCI grant number 5U01CA177799, Saving lives at birth and USP/PQM cooperative agreement. ZBW is supported by NIGMS Training Program in Biomolecular Pharmacology T32GM008541.}

\showacknow 


\bibliographystyle{pnas2011}
\bibliography{bioaptamers}

\end{document}